\documentclass[twocolumn,english,prl,floatfix,superscriptaddress]{revtex4-1}
\usepackage{mathptmx}
\usepackage[T1]{fontenc}
\usepackage[latin9]{inputenc}
\usepackage{color}
\usepackage{babel}
\usepackage{float}
\usepackage{booktabs}
\usepackage{amsmath}
\usepackage{amssymb}
\usepackage{graphicx}
\usepackage{epstopdf}
\PassOptionsToPackage{normalem}{ulem}
\usepackage{ulem}
\usepackage[unicode=true,pdfusetitle,
 bookmarks=true,bookmarksnumbered=false,bookmarksopen=false,
 breaklinks=false,pdfborder={0 0 0},pdfborderstyle={},backref=false,colorlinks=true]
 {hyperref}
\hypersetup{
 pdfborderstyle={},pdfborderstyle={},allcolors=magenta}

\makeatletter

\providecommand{\tabularnewline}{\\}

\usepackage{babel}
\usepackage{babel}
\usepackage{babel}
\usepackage{babel}
\usepackage{times}
\usepackage{pxfonts}
\usepackage{color}
\usepackage{braket}
\usepackage{dsfont}

\makeatother

\begin{document}

\title{Direct estimation of quantum coherence by collective measurements}

\author{Yuan Yuan}

\affiliation{CAS Key Laboratory of Quantum Information, University of Science
and Technology of China, Hefei, 230026, China}

\affiliation{Department of Physics, East China University of Science and Technology, Shanghai, 200237, China}

\affiliation{Synergetic Innovation Center of Quantum Information and Quantum Physics,
University of Science and Technology of China, Hefei, Anhui 230026, China}

\author{Zhibo Hou}

\affiliation{CAS Key Laboratory of Quantum Information, University of Science
and Technology of China, Hefei, 230026, China}

\affiliation{Synergetic Innovation Center of Quantum Information and Quantum Physics,
University of Science and Technology of China, Hefei, Anhui 230026,
China}

\author{Jun-Feng Tang}

\affiliation{CAS Key Laboratory of Quantum Information, University of Science
and Technology of China, Hefei, 230026, China}

\affiliation{Synergetic Innovation Center of Quantum Information and Quantum Physics,
University of Science and Technology of China, Hefei, Anhui 230026,
China}

\author{Alexander Streltsov}
\email{streltsov.physics@gmail.com}

\affiliation{Centre for Quantum Optical Technologies IRAU, Centre of New Technologies,
University of Warsaw, Banacha 2c, 02-097 Warsaw, Poland}

\author{Guo-Yong Xiang}
\email{gyxiang@ustc.edu.cn}

\selectlanguage{english}%

\affiliation{CAS Key Laboratory of Quantum Information, University of Science
and Technology of China, Hefei, 230026, China}

\affiliation{Synergetic Innovation Center of Quantum Information and Quantum Physics,
University of Science and Technology of China, Hefei, Anhui 230026,
China}

\author{Chuan-Feng Li}

\affiliation{CAS Key Laboratory of Quantum Information, University of Science
and Technology of China, Hefei, 230026, China}

\affiliation{Synergetic Innovation Center of Quantum Information and Quantum Physics,
University of Science and Technology of China, Hefei, Anhui 230026,
China}

\author{Guang-Can Guo}

\affiliation{CAS Key Laboratory of Quantum Information, University of Science
and Technology of China, Hefei, 230026, China}

\affiliation{Synergetic Innovation Center of Quantum Information and Quantum Physics,
University of Science and Technology of China, Hefei, Anhui 230026,
China}

\begin{abstract}
The recently established resource theory of quantum coherence allows
for a quantitative understanding of the superposition principle, with
applications reaching from quantum computing to quantum biology. While
different quantifiers of coherence have been proposed in the literature,
their efficient estimation in today's experiments remains a challenge.
Here, we introduce a collective measurement scheme for estimating
the amount of coherence in quantum states, which requires entangled
measurements on two copies of the state. As we show by numerical simulations,
our scheme outperforms other estimation methods based on tomography
or adaptive measurements, leading to a higher precision in a large
parameter range for estimating established coherence quantifiers of
qubit and qutrit states. We show that our method is accessible with
today's technology by implementing it experimentally with photons,
finding a good agreement between experiment and theory.
\end{abstract}
\maketitle

\noindent\textbf{Introduction}\\
Quantum coherence is the most distinguished
feature of quantum mechanics, characterizing the superposition properties
of quantum states. An operational resource theory of coherence has
been established in the last years~\cite{Baumgratz,coherence1,coherence2,coherence3,coherence4,review},
allowing for a systematic study of quantum coherence in quantum technology
\cite{review}, including quantum algorithms \cite{algorithm1,algorithm2},
quantum computation~\cite{CoherenceComputation}, quantum key distribution~\cite{QKD},
quantum channel discrimination \cite{discrimination1,discrimination2},
and quantum metrology \cite{metrology1,metrology2,metrology3}. Moreover,
quantum coherence is closely related to other quantum resources, such
as asymmetry \cite{asymmetry1,asymmetry2}, entanglement~\cite{CoherenceEntanglement1,CoherenceEntanglement2}
and other quantum correlations \cite{relation1}; the manipulation
of coherence and conversion between coherence and quantum correlations
in bipartite and multipartite systems has been explored both theoretically~\cite{conversion1,conversion2,conversion3,conversion4}
and experimentally~\cite{kangda1,kangda2}. Highly relevant
from the experimental perspective is the recent progress on coherence
theory in the finite copy regime, in particular regarding single-shot
coherence distillation~\cite{SingleShotDistillation1,SingleShotDistillation2,SingleShotDistillation3,SingleShotDistillation4},
coherence dilution~\cite{SingleShotDilution}, and incoherent state
conversion~\cite{kangda3}. Being a fundamental property of quantum
systems, coherence plays an important role in quantum thermodynamics~\cite{thermo1,thermo2,thermo3,thermo4,thermo5,thermo6,thermo7},
nanoscale physics~\cite{nano}, transport theory~\cite{transport1,transport2},
biological systems~\cite{biology1,biology2,biology3,biology4,biology5,biology6},
and for the study of the wave-particle duality \cite{wp1,wp2,wp3}.

Having identified quantum coherence as a valuable feature of quantum
systems, it is important to develop methods for its rigorous quantification.
First attempts for resource quantification were made in the resource
theory of entanglement~\cite{QuantifyingEntanglement,ReviewEntanglement},
leading to various entanglement measures based on physical or mathematical
considerations. The common feature of all resource quantifiers is
the postulate that they should not increase under \emph{free operations}
of the theory, which in entanglement theory are known as ``local
operations and classical communication''. In the resource theory
of coherence, the free operations are \emph{incoherent operations},
corresponding to quantum measurements which cannot create coherence
for individual measurement outcomes~\cite{Baumgratz}.

While various coherence measures have been proposed~\cite{review},
an important issue is how to efficiently estimate them in experiments.
Clearly, one possibility is to perform quantum state tomography~\cite{tomography}
and then use the derived state density matrix to estimate the amount
of coherence. However, estimation of coherence measures in general
does not require the complete information about the state of the system,
a fact which has been exploited in various approaches for detecting
and estimating coherence of unknown quantum states~\cite{skew,wang,tong,njp}.

In this paper, we put forward a general method to directly measure
quantum coherence of an unknown quantum state using two-copy collective
measurement scheme (CMS) \cite{coll1,coll2,coll3,hou}. We simulate
the performance of this method for qubits and qutrits and compare
the precision of CMS with other methods for coherence estimation,
including tomography. The simulations show that in certain setups
CMS outperforms all other schemes discussed in this work. We also
report an experimental demonstration of CMS for qubit states. The
collective measurements are performed on two identically prepared
qubits which are encoded in two degrees of freedom of a single photon.
In this way, we experimentally obtain two widely studied coherence
measures, finding a good agreement between the numerical simulations
and the experimental results.

\medskip{}

\noindent\textbf{Theoretical framework}\\
We aim to estimate coherence of a quantum state $\rho$ by performing measurements
on two copies of the state. As quantifiers of coherence we use the
$\ell_{1}$-norm of coherence $C_{\ell_{1}}$ and the relative entropy
of coherence $C_{r}$, defined as~\cite{Baumgratz}
\begin{align}
C_{\ell_{1}}(\rho) & =\sum_{i\neq j}\left|\rho_{ij}\right|,\\
C_{r}(\rho) & =S(\rho_{diag})-S(\rho).
\end{align}
Here, $S(\rho)=-\mathrm{Tr}[\rho\log_{2}\rho]$ is the von Neumann
entropy, $\rho_{diag}=\sum_{i}|i\rangle\langle i|\rho|i\rangle\langle i|$,
and we consider coherence with respect to the computational basis
$\{\ket{i}\}$. For single-qubit states with Bloch vector $\boldsymbol{r}=(r_{x},r_{y},r_{z})$,
both quantities can be expressed as~\cite{Hu}
\begin{align}
C_{\ell_{1}}(\rho) & =\sqrt{r_{x}^{2}+r_{y}^{2}},\label{eq:Cl1Qubit}\\
C_{r}(\rho) & =h\left(\frac{1+\left|r_{z}\right|}{2}\right)-h\left(\frac{1+r}{2}\right)\label{eq:CrQubit}
\end{align}
with the binary entropy $h(x)=-x\log_{2}x-(1-x)\log_{2}(1-x)$ and
the Bloch vector length $r=(r_{x}^{2}+r_{y}^{2}+r_{z}^{2})^{1/2}$.

\begin{figure}
\includegraphics[width=8cm,height=12cm]{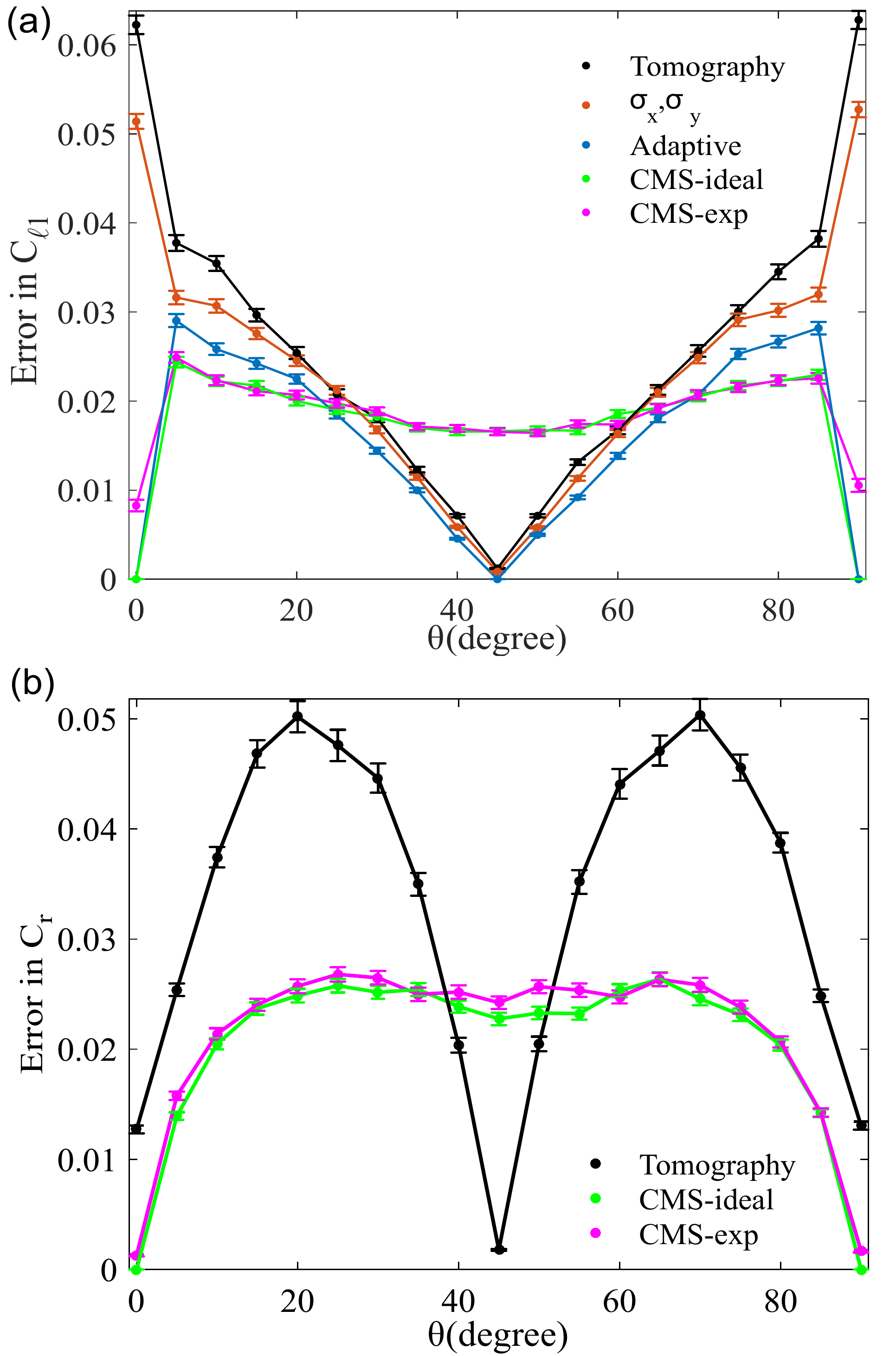} \caption{\label{fig:QubitResults1}Mean errors for estimating $\ell_{1}$-norm
coherence in (a) and relative entropy coherence in (b) for a family
of qubit states with different measurement schemes. The states have
the form $|\Psi\rangle=\sin\theta|0\rangle+\cos\theta|1\rangle$ with
$\theta$ ranging from $0$ to $\pi/2$. In (a), the performances
of CMS (numerical simulation and experiment); $\sigma_{x}$, $\sigma_{y}$
measurement (simulation); two-step adaptive measurements (simulation);
and tomography (simulation) are shown for comparison. In (b), the
performances of CMS (numerical simulation and experiment) and tomography
(simulation) are shown for comparison. The sample size is $N=1200$.
Each data point is the average of $1000$ repetitions, and the error
bar denotes the standard deviation.}
\end{figure}

\begin{figure}
\includegraphics[width=8.7cm,height=5.6cm]{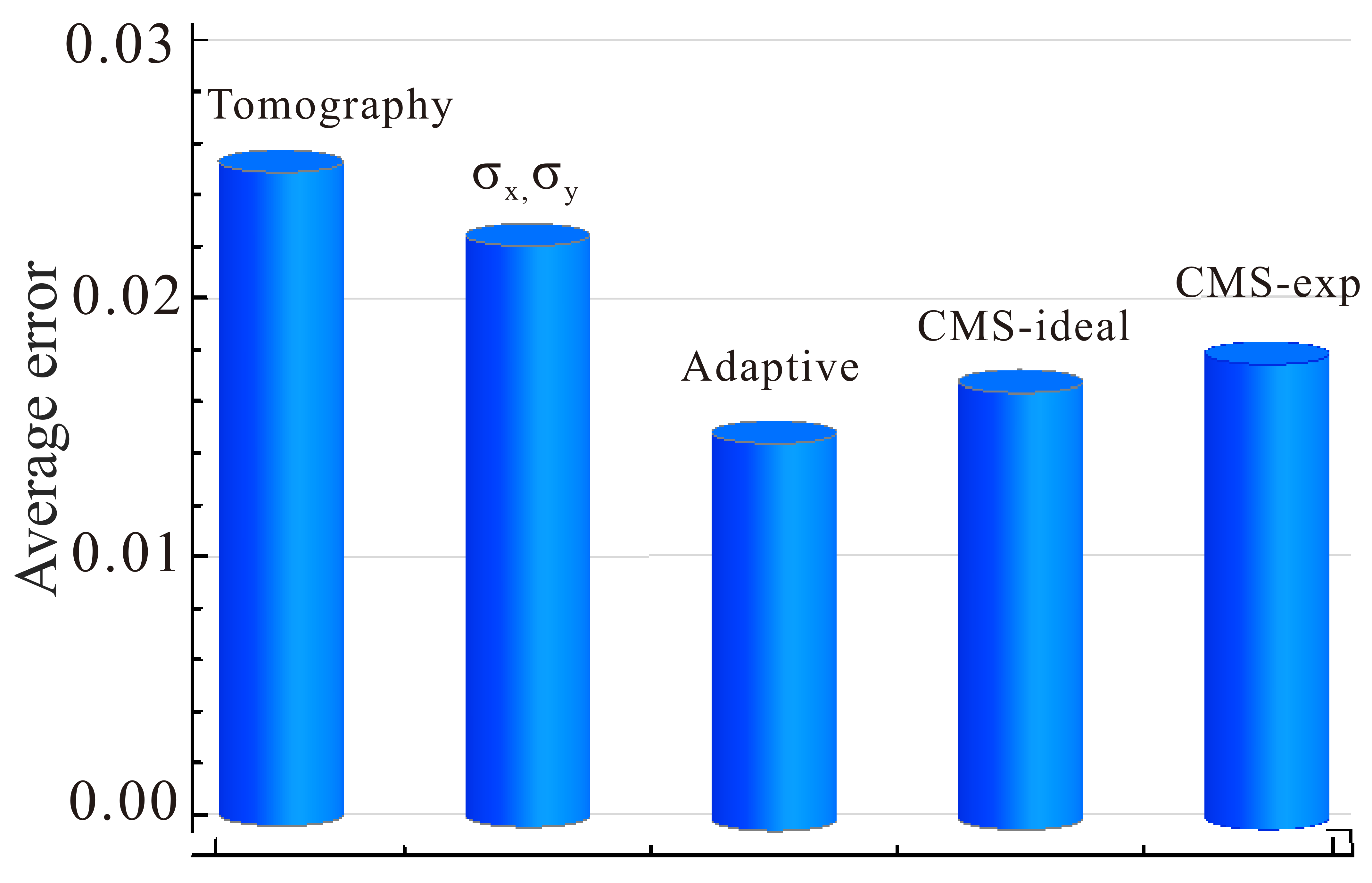} \caption{\label{fig:QubitResults2}Average results of the mean error for all
input states shown in Fig.~\ref{fig:QubitResults1}(a). The corresponding
average values of these methods from left to right: $0.0263$, $0.0234$,
$0.0156$, $0.0176$, $0.0187$.}
\end{figure}

In the next step, we will express both $C_{\ell_{1}}$ and $C_{r}$
in terms of the outcome probabilities of a collective measurement
in the maximally entangled basis, performed on two copies $\rho\otimes\rho$.
We denote the corresponding outcome probabilities as $P_{i}=\mathrm{Tr}[M_{i}\rho\otimes\rho]$,
where
\begin{equation}
\begin{aligned}M_{1} & =\ket{\psi^{+}}\!\bra{\psi^{+}},\,\,\,\,\,M_{2}=\ket{\psi^{-}}\!\bra{\psi^{-}},\\
M_{3} & =\ket{\varphi^{+}}\!\bra{\varphi^{+}},\,\,\,\,\,M_{4}=\ket{\varphi^{-}}\!\bra{\varphi^{-}}
\end{aligned}
\label{eq:Mi}
\end{equation}
are projectors onto maximally entangled states $|\psi^{\pm}\rangle=(|01\rangle\pm|10\rangle)/\sqrt{2}$
and $\ket{\varphi^{\pm}}=(\ket{00}\pm\ket{11})/\sqrt{2}$. As
we show in Supplement~\ref{sec:qubit}, the outcome
probabilities fulfill the relations\begin{subequations}\label{eq:two-copy-2}
\begin{align}
r_{x}^{2}+r_{y}^{2} & =2(P_{1}-P_{2}),\\
|r_{z}| & =\sqrt{2\left(P_{3}+P_{4}\right)-1},\\
r & =\sqrt{1-4P_{2}}.
\end{align}
\end{subequations}Thus, both coherence measures $C_{\ell_{1}}$ and
$C_{r}$ can be expressed as simple functions of $P_{i}$. We further
note that in general CMS can be used to estimate absolute values of
the Bloch vector components of a single-qubit state $\rho$. This
implies that CMS allows to evaluate any coherence measure of single-qubit
states, as any such measure is a function of the absolute values of
the Bloch coordinates, see Supplement~\ref{sec:qubit} for more details.

In the following, we use numerical simulation to compare the collective
measurement scheme (CMS) discussed above to three alternative schemes
for measuring $C_{\ell_{1}}$ for single-qubit states. The first alternative
scheme is to directly measure observables $\sigma_{x}$ and $\sigma_{y}$,
and estimate $C_{\ell_{1}}$ via Eq.~(\ref{eq:Cl1Qubit}). The second
scheme is a two-step adaptive measurement: step one is to measure
observables $\sigma_{x}$ and $\sigma_{y}$; based on the feedback
results of the first step, step two is to choose optimal observable
$a\sigma_{x}+b\sigma_{y}$ to obtain $|\!\braket{0|\rho|1}\!|$. The
third alternative scheme is to perform state tomography and then,
subject to the derived density matrix, to estimate the value of $C_{\ell_{1}}$.
We further use the tomography results to estimate the relative entropy
of coherence $C_{r}$ via Eq.~(\ref{eq:CrQubit}), and compare
the performance of the estimation with CMS.

For the numerical simulation we use single-qubit states
\begin{equation}
\ket{\Psi}=\sin\theta\ket{0}+\cos\theta\ket{1}\label{eq:Psi}
\end{equation}
with $\theta$ ranging from $0$ to $\pi/2$. All simulations are
performed on $N=1200$ copies of $\ket{\Psi}$. We further repeat
each simulation $1000$ times and average the numerical data over
all repetitions. We are in particular interested in the error of the
estimation:
\begin{equation}
\varepsilon=\left|C_{\mathrm{est}}(\rho)-C(\rho)\right|,
\end{equation}
where $C_{\mathrm{est}}$ and $C$ are the estimated and the actual
coherence measures, respectively. Fig.~\ref{fig:QubitResults1}(a)
shows the results of numerical simulation for $C_{\ell_{1}}$,
together with experimental data; the experiment will be discussed
in more detail below. Each data point in the figure is the average
of $T=1000$ repetitions, i.e., $\frac{1}{T}\sum_{i=1}^{T}\varepsilon_{i}$,
where $\varepsilon_{i}$ is the error of the $i$th measurement. The
error bar denotes the standard deviation of $\varepsilon_{i}$. Fig.~\ref{fig:QubitResults1}(b)
shows the corresponding comparison between CMS and tomography for
estimating the relative entropy of coherence $C_{r}$.

As we see from the data shown in Fig.~\ref{fig:QubitResults1}(a,b),
there is a range of $\theta$ where CMS outperforms all other schemes,
leading to the smallest error. Moreover, while the error in general
depends on $\theta$, this dependence is comparably weak for CMS.
To compare the accuracy achieved by different estimation methods more
intuitively and clearly, we average the mean error for all input states
shown in Fig.~\ref{fig:QubitResults1}(a), and the average results
are shown in Fig.~\ref{fig:QubitResults2}. For the estimation
of $C_{\ell_{1}}$ the adaptive measurement scheme outperforms CMS
on average, which again outperforms all other estimation schemes presented
above. In Supplement~\ref{sec:Cf} we further report theoretical
and experimental results for estimating coherence of formation~\cite{formation,coherence2}
for qubits. Also in this case CMS outperforms all other schemes discussed
in this paper in a certain range of $\theta$.

While the above discussion was restricted to qubit systems, the CMS
method can also be applied to estimate $C_{\ell_{1}}$ for states
of higher dimensions. We consider an arbitrary quantum state $\rho=\sum_{i,j}\rho_{ij}|i\rangle\langle j|$,
where $i,j=0,1,...,d-1$ and $d$ is the dimension of Hilbert space.
After an appropriate set of collective measurements are performed
on the two-copy state $\rho\otimes\rho$, we find that the absolute
value of the off-diagonal element $|\rho_{ij}|$ for $i\neq j$ can
be expressed as
\begin{equation}
\left|\rho_{ij}\right|=\sqrt{\frac{1}{2}\left(\mathrm{Tr}\left[\rho^{\otimes2}\ket{\psi_{ij}^{+}}\!\bra{\psi_{ij}^{+}}\right]-\mathrm{Tr}\left[\rho^{\otimes2}\ket{\psi_{ij}^{-}}\!\bra{\psi_{ij}^{-}}\right]\right)},\label{eq:pij}
\end{equation}
where $|\psi_{ij}^{\pm}\rangle=(|ij\rangle\pm|ji\rangle)/\sqrt{2}$.
Therefore, the $\ell_{1}$-norm coherence can be written as
\begin{equation}
C_{\ell_{1}}(\rho)=2\sum\limits _{j>i}\sqrt{\frac{1}{2}\left(\mathrm{Tr}\left[\rho^{\otimes2}\ket{\psi_{ij}^{+}}\!\bra{\psi_{ij}^{+}}\right]-\mathrm{Tr}\left[\rho^{\otimes2}\ket{\psi_{ij}^{-}}\!\bra{\psi_{ij}^{-}}\right]\right)}.\label{C}
\end{equation}

\begin{figure}
\includegraphics[width=1\columnwidth]{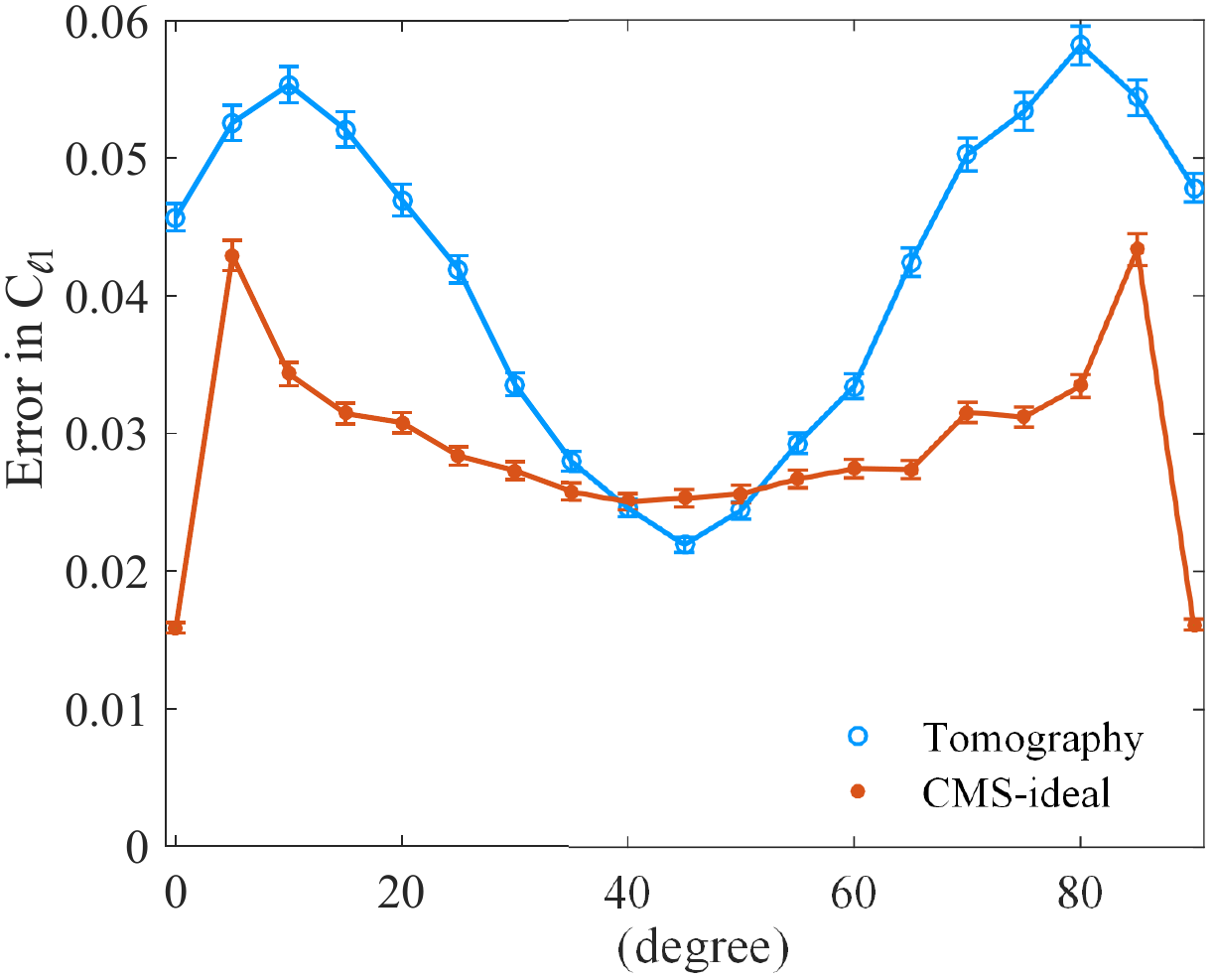} \caption{\label{fig:Qutrit}Mean error for estimating $C_{\ell_{1}}$ for a
family of qutrit states in Eq.~(\ref{eq:Qutrit}) for CMS and qutrit
state tomography (numerical simulation). The sample size
is $N=1200$. Each data point is the average of $1000$ repetitions,
and the error bars denote the standard deviation.}
\end{figure}

We use numerical simulation to compare the performance of the CMS
method to the qutrit state tomography (see Supplement~\ref{sec:Qutrit-state-tomography})
for the family of qutrit states
\begin{equation}
|\Phi\rangle=\frac{1}{\sqrt{2}}(\sin\alpha|0\rangle+\cos\alpha|1\rangle+|2\rangle),\label{eq:Qutrit}
\end{equation}
with $\alpha$ ranging from $0$ to $\pi/2$. The results of the simulation
are shown in Fig.~\ref{fig:Qutrit}. As before, we use $N=1200$
copies of the state $\ket{\Phi}$ for both CMS and state tomography,
and average over $1000$ repetitions. The results show that CMS outperforms
the tomography method for a large range of $\alpha$. Apart from a
higher accuracy, the CMS method requires only a single measurement
setup, while four measurement setups are required for qutrit tomography.

\medskip{}

\begin{figure}
\includegraphics[width=1\columnwidth]{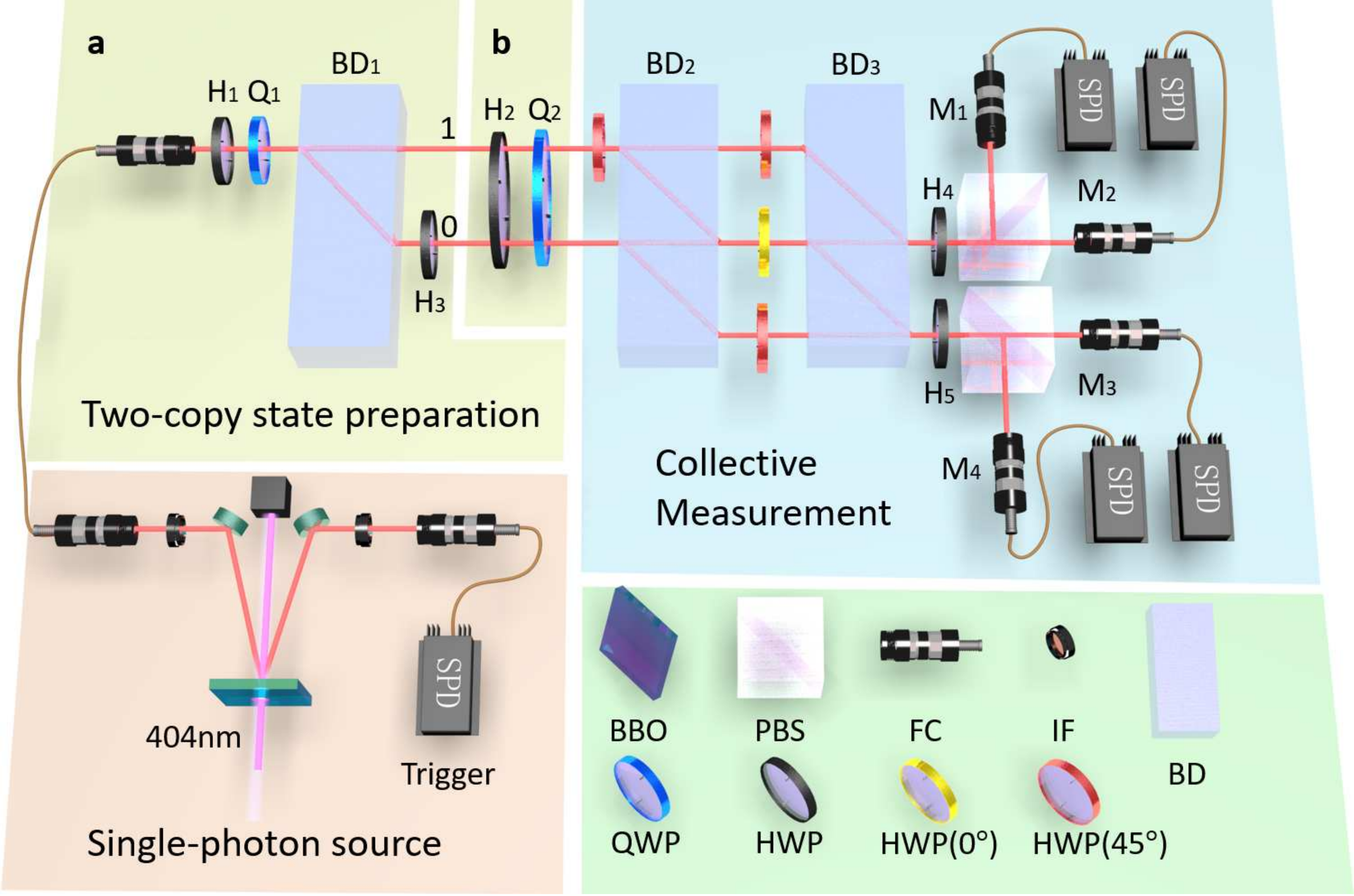} \caption{\label{fig:Experimental-setup}Experimental setup for measuring coherence
of qubit with collective measurements. The setup consists of three
modules designed for single-photon source, two-copy state preparation
(\textbf{a}, \textbf{b}) and collective measurement, respectively.
In the single-photon source module, the photon pairs generated in
spontaneous parametric down-conversion are coupled into single-mode
fibers separately. One photon is detected by a single-photon detector
(SPD) acting as a trigger. In the two-copy state preparation module,
(\textbf{a}) prepares the first copy in the path degree of freedom
of the photon; (\textbf{b}) prepares the second copy in the polarization
degree of freedom of the photon. In the collective measurement module,
combinations of beam displacers (BDs) and half wave plates (HWPs)
with certain angular settings are used to realize collective measurement,
where H$_{4}$ and H$_{5}$ are $22.5^{\circ}$. Four SPDs $M_{1}$
to $M_{4}$ correspond to the four outcomes of the collective measurement.}
\end{figure}

\noindent \textbf{Experimental implementation}\\
The experimental setup for realizing CMS to estimate coherence of qubit states is presented
in Fig.~\ref{fig:Experimental-setup}. The setup is composed of three
modules designed for single-photon source, two-copy state preparation,
and collective measurements, respectively. In the single-photon source
module, a 80-mW cw laser with a 404-nm wavelength (linewidth=5 MHz)
pumps a type-II beamlike phase-matching beta-barium-borate (BBO, 6.0$\times$6.0$\times$2.0
~mm$^{3}$, $\theta=40.98^{\circ}$) crystal to produce a pair of
photons with wavelength $\lambda=808$~nm. The two photons pass through
two interference filters (IF) whose FWHM (full width at half maximum)
is $3$ nm. The photon pairs generated in spontaneous parametric down-conversion
(SPDC) are coupled into single-mode fibers separately. One photon
is detected by a single-photon detector acting as a trigger. In the
two-copy state preparation module, we first prepare copy $1$ in the
path degree of freedom of single photon, i.e., the first qubit encoded
in positions $1$ and $0$ (see \textbf{a} in Fig.~\ref{fig:Experimental-setup}).
After passing a half-wave plate (HWP) and a quarter-wave plate (QWP)
with deviation angles H$_{1}$, Q$_{1}$, the photon is prepared in
the desired state $\rho$. To encode the polarization state into the
path degree of freedom, beam displacer (BD$_{1}$) is used to displace
the horizontal polarization (H) component into path $0$, which is
4-mm away from the vertical polarization (V) component in path $1$;
then a HWP (H$_{3}$) with deviation angle $45^{\circ}$ is placed
in path $0$. The resulting photon is described by the state $\rho\otimes|V\rangle\langle V|$.
Then we encode the second copy of $\rho$ into the polarization degree
of freedom of single photon using a HWP and a QWP with deviation angles
H$_{2}$, Q$_{2}$ (see \textbf{b} in Fig.~\ref{fig:Experimental-setup}).
In this way, we can prepare the desired two-copy state $\rho\otimes\rho$.

%\begin{figure}
%\includegraphics[width=1\columnwidth]{bell} \caption{\label{fig:Experimental-results}Experimental verification of the
%collective measurement realized. Here $|\psi^{+}\rangle$, $|\psi^{-}\rangle$,
%$|\varphi^{+}\rangle$ and $|\varphi^{-}\rangle$ denote the four
%input states. Each input state is prepared and measured $5000$ times.
%The frequencies of obtaining the four outcomes are plotted using different
%colors; here the error bars are too small to be visible. For comparison,
%the probabilities in the ideal scenario are plotted in gray shadow.}
%\end{figure}

The collective measurement module realizes a measurement on $\rho\otimes\rho$
in the maximally entangled basis, where $M_{i}$ are given in Eq.~(\ref{eq:Mi}).
When estimating the $\ell_{1}$-norm coherence, only the probabilities
of the outcomes $M_{1}$ and $M_{2}$ are used, see the discussion
below Eq.~(\ref{eq:Cl1Qubit}). The probabilities of all outcomes
are used for estimating the relative entropy of coherence, see Eq.~(\ref{eq:CrQubit}).
To verify the experimental implementation of the collective measurement,
we take the conventional method of measuring the probability distributions
after preparing the input states $|\psi^{+}\rangle$, $|\psi^{-}\rangle$,
$|\varphi^{+}\rangle$ and $|\varphi^{-}\rangle$. These input states
can be prepared by choosing proper rotation angles H$_{1}$, Q$_{1}$,
H$_{2}$, H$_{3}$ as specified in Supplement~\ref{sec:Angles}. Each input state
is prepared and measured $5000$ times, and the probability of obtaining the
outcomes $M_{1}$, $M_{2}$, $M_{3}$ and $M_{4}$ are $0.9981\pm0.0006$, $0.9973\pm0.0007$,
$0.9962\pm0.0009$ and $0.9961\pm0.0009$, respectively (ideal value is 1). The theoretical
values of other probability distributions for the input states are all $0$, experimentally the maximum
error of other probability is $0.0037\pm0.0009$.

The experimental deterministic realization of the collective measurement
allows us to estimate the amount of coherence with a single measurement
setup. We experimentally investigate the error achieved by CMS when
the input states $\ket{\Psi}$ have the form~(\ref{eq:Psi}) with
$\theta$ ranging from $0$ to $\pi/2$. The sample size of the experiment
is $N=1200$ copies of $\ket{\Psi}$; same sample size has been used
in the numerical simulations reported above. As in the numerical simulation,
we average over $1000$ repetitions of the experiment. The experimental results for the estimation precision
of $C_{\ell_{1}}$ and $C_{r}$ are shown in Fig.~\ref{fig:QubitResults1}(a)
and Fig.~\ref{fig:QubitResults1}(b), respectively. The experimental
data is in good agreement with the theoretical prediction.

\medskip{}

\noindent \textbf{Discussion}\\
%\textbf{\emph{Conclusions.}}
We introduce a general method to directly
measure quantum coherence of an unknown quantum state using two-copy
collective measurement, focusing on two established coherence quantifiers:
$\ell_{1}$-norm coherence and relative entropy coherence. As we demonstrate
by numerical simulation for qubit and qutrit states, in a certain
parameter region the collective measurement scheme outperforms other
estimation techniques, including methods based on adaptive $\sigma_{x}$,
$\sigma_{y}$ measurement for qubits, and tomography-based coherence
estimation for qubits and qutrits. We test our results by experimentally
estimating the $\ell_{1}$-norm coherence and relative entropy coherence
of qubit states by collective measurements in optical setup, finding
good agreement between theory and experiment. For single-qubit states
our method allows to estimate absolute values of the Bloch coordinates,
implying that any coherence quantifier of a qubit can be estimated
with the collective measurement scheme.

Although the precision achieved by our method is not always better
than by adaptive measurement, our scheme has several advantages with
respect to other techniques. In particular, our method does not need
any optimization procedures or feedback, which are required for coherence
estimation via adaptive measurements. Moreover, the entire experiment
can be performed in a single measurement setup. Thus, our work provides
a simple method to measure coherence, and highlights the application
of collective measurement in quantum information processing.
\medskip{}

\bigskip
\noindent\textbf{Acknowledgements}\\
We thank Huangjun Zhu for fruitful discussions. This work is supported by the National Natural Science Foundation of China (Grants No. 11574291 and No. 11774334),
National Key Research and Development Program of China (Grants No.
2016YFA0301700 and No.2017YFA0304100), and Anhui Initiative in Quantum
Information Technologies. AS acknowledges financial support by the
``Quantum Optical Technologies'' project, carried out within the
International Research Agendas programme of the Foundation for Polish
Science co-financed by the European Union under the European Regional
Development Fund.

 \newpage

 %%%%%%%%% Merge with supplemental materials %%%%%%%%%%
 %%%%%%%%% Prefix a "S" to all equations, figures, tables and reset the counter
 %%%%%%%%%
 \setcounter{equation}{0}
 \setcounter{figure}{0}
 \setcounter{table}{0}

 \makeatletter
 \renewcommand{\theequation}{S\arabic{equation}}
 \renewcommand{\thefigure}{S\arabic{figure}}
 \renewcommand{\thetable}{S\arabic{table}}

\onecolumngrid

\begin{center}
 \textbf{\large Direct estimation of quantum coherence by collective measurements: Supplement}
\end{center}

\twocolumngrid

\subsection{\label{sec:qubit}Estimating general coherence measures for qubits with collective measurements}

For a single-qubit state $\rho$ with Bloch vector $\boldsymbol{r}=(r_{x},r_{y},r_{z})$
the probabilities $P_{i}=\mathrm{Tr}[M_{i}\rho\otimes\rho]$ are given
explicitly as \begin{subequations}
\begin{align}
P_{1} & =\braket{\psi^{+}|\rho\otimes\rho|\psi^{+}}=\frac{1}{4}\left(1+r_{x}^{2}+r_{y}^{2}-r_{z}^{2}\right),\\
P_{2} & =\braket{\psi^{-}|\rho\otimes\rho|\psi^{-}}=\frac{1}{4}\left(1-r_{x}^{2}-r_{y}^{2}-r_{z}^{2}\right),\\
P_{3} & =\braket{\phi^{+}|\rho\otimes\rho|\phi^{+}}=\frac{1}{4}\left(1+r_{x}^{2}-r_{y}^{2}+r_{z}^{2}\right),\\
P_{4} & =\braket{\phi^{-}|\rho\otimes\rho|\phi^{-}}=\frac{1}{4}\left(1-r_{x}^{2}+r_{y}^{2}+r_{z}^{2}\right).
\end{align}
\end{subequations}It thus follows that collective measurements
can be used to evaluate absolute values of the Bloch coordinates:\begin{subequations}
\begin{align}
\left|r_{x}\right| & =\sqrt{2\left(P_{1}+P_{3}\right)-1},\\
\left|r_{y}\right| & =\sqrt{2\left(P_{1}+P_{4}\right)-1},\\
\left|r_{z}\right| & =\sqrt{2\left(P_{3}+P_{4}\right)-1}.
\end{align}
\end{subequations} From these results, it is straightforward
to verify Eqs.~(6) of the main text.

In the following, $C(r_{x},r_{y},r_{z})$ will denote a coherence
measure for a qubit state $\rho$ with Bloch vector $\boldsymbol{r}=(r_{x},r_{y},r_{z})$.
As we will now show, for single-qubit states any coherence measure
$C$ depends only on the absolute values of the Bloch vector coordinates.
For this, it is enough to show that
\begin{equation}
C(r_{x},r_{y},r_{z})=C(-r_{x},r_{y},r_{z})=C(r_{x},-r_{y},r_{z})=C(r_{x},r_{y},-r_{z})\label{eq:QubitCoherence}
\end{equation}
for any coherence measure $C$ and any Bloch vector. This can be seen
by noting that the vector $(r_{x},r_{y},r_{z})$ can be transformed
into the vector $(-r_{x},r_{y},r_{z})$ via a rotation around the
$z$-axis, which corresponds to an incoherent unitary operation. Since
any coherence measure is invariant under incoherent unitaries, it
follows that $C(r_{x},r_{y},r_{z})=C(-r_{x},r_{y},r_{z})$. By similar
arguments we obtain $C(r_{x},r_{y},r_{z})=C(r_{x},-r_{y},r_{z})$.
Moreover, note that $\sigma_{x}$ is an incoherent unitary inducing
the transformation $(r_{x},r_{y},r_{z})\rightarrow(r_{x},-r_{y},-r_{z})$,
and thus it must be that $C(r_{x},r_{y},r_{z})=C(r_{x},-r_{y},-r_{z})$.
Combining these arguments completes the proof of Eq.~(\ref{eq:QubitCoherence}).

\subsection{\label{sec:Cf}Estimating coherence of formation for qubits}
\begin{figure}[htp]
\includegraphics[width=8.8cm,height=11.9cm]{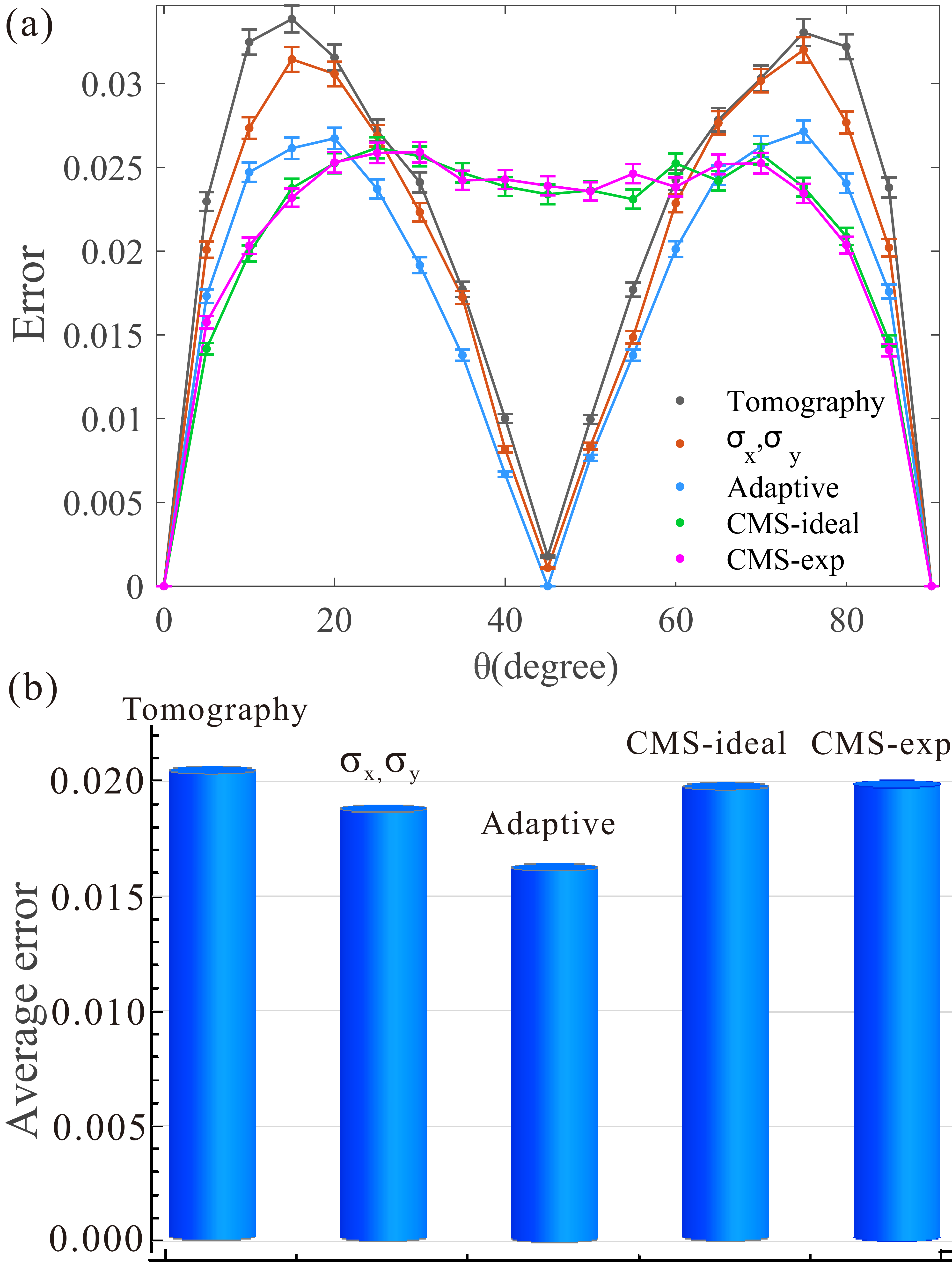}
\caption{\label{fig:Cf}(a) Mean errors for estimating coherence of formation
for a family of qubit states with different measurement schemes. The
states have the form $|\Psi\rangle=\sin\theta|0\rangle+\cos\theta|1\rangle$
with $\theta$ ranging from $0$ to $\pi/2$. In (a), the performances
of CMS (numerical simulation and experiment); $\sigma_{x}$, $\sigma_{y}$
measurement (simulation); two-step adaptive measurements (simulation);
and tomography (simulation) are shown for comparison. The sample size
is $N=1200$. Each data point is the average of $1000$ repetitions,
and the error bar denotes the standard deviation. (b) Average results
of the mean error for all input states shown in (a). The corresponding
average values of these methods are from left to right: 0.0211, 0.0194,
0.0168, 0.0204, 0.0205.}
\end{figure}

We will now apply the collective measurement scheme to estimate the
coherence of formation, which for single-qubit states takes the form~\cite{formation}
\begin{equation}
C_{f}(\rho)=h\left(\frac{1+\sqrt{1-4|\rho_{01}|^{2}}}{2}\right),
\end{equation}
where $h(x)=-x\log_{2}x-(1-x)\log_{2}(1-x)$ is the binary entropy.
We perform numerical simulations and an optical experiment, following
the same procedure as for estimating $C_{\ell_{1}}$. In particular,
we use $N=1200$ copies of the state
\begin{equation}\label{qubit}
|\Psi\rangle=\sin\theta|0\rangle+\cos\theta|1\rangle
\end{equation}
with $\theta\in[0,\pi/2]$, and average the data over $1000$ repetitions.
The results of the numerical simulation and experiment are shown in
Fig.~\ref{fig:Cf}. Also in this case we compare the numerical simulation
of CMS with three alternative schemes for coherence estimation, see
main text for details.

\subsection{\label{sec:Qutrit-state-tomography}Qutrit state tomography}

The density matrix of an unknown qutrit state is reconstructed by
performing projective measurements in 4 mutually unbiased bases. In
the following, vectors $\ket{\xi_{ij}}$ form an orthonormal basis
for a given $i$, i.e., $\braket{\xi_{ij}|\xi_{ik}}=\delta_{jk}$,
and are mutually unbiased for different $i$: $|\!\braket{\xi_{ij}|\xi_{kl}}\!|=1/\sqrt{3}$
for $i\neq k$. The vectors can be explicitly given as follows:
\begin{equation}
\begin{aligned} & |\xi_{00}\rangle=|0\rangle,|\xi_{01}\rangle=|1\rangle,|\xi_{02}\rangle=|2\rangle,\\
 & |\xi_{10}\rangle=\frac{1}{\sqrt{3}}(|0\rangle+|1\rangle+|2\rangle),\\
 & |\xi_{11}\rangle=\frac{1}{\sqrt{3}}(|0\rangle+e^{\frac{2i\pi}{3}}|1\rangle+e^{-\frac{2i\pi}{3}}|2\rangle),\\
 & |\xi_{12}\rangle=\frac{1}{\sqrt{3}}(|0\rangle+e^{-\frac{2i\pi}{3}}|1\rangle+e^{\frac{2i\pi}{3}}|2\rangle),\\
 & |\xi_{20}\rangle=\frac{1}{\sqrt{3}}(e^{\frac{2i\pi}{3}}|0\rangle+|1\rangle+|2\rangle),\\
 & |\xi_{21}\rangle=\frac{1}{\sqrt{3}}(|0\rangle+e^{\frac{2i\pi}{3}}|1\rangle+|2\rangle),\\
 & |\xi_{22}\rangle=\frac{1}{\sqrt{3}}(|0\rangle+|1\rangle+e^{\frac{2i\pi}{3}}|2\rangle),\\
 & |\xi_{30}\rangle=\frac{1}{\sqrt{3}}(e^{-\frac{2i\pi}{3}}|0\rangle+|1\rangle+|2\rangle),\\
 & |\xi_{31}\rangle=\frac{1}{\sqrt{3}}(|0\rangle+e^{-\frac{2i\pi}{3}}|1\rangle+|2\rangle),\\
 & |\xi_{32}\rangle=\frac{1}{\sqrt{3}}(|0\rangle+|1\rangle+e^{-\frac{2i\pi}{3}}|2\rangle).
\end{aligned}
\end{equation}
For numerical simulation of coherence estimation via qutrit state
tomography reported in the main text (see Fig.~\ref{fig:Qutrit}),
we use the sample size $N=1200$, and repeat the procedure 1000 times
to infer the state from the results of the measurement. Here we use
the method of maximum-likelihood reconstruction presented in Ref.
\cite{MLE}.

\subsection{\label{sec:Angles}Angles of wave plates used to prepare the states}

In the state preparation module, H$_{1}$, H$_{2}$, H$_{3}$, Q$_{1}$,
Q$_{2}$ are the rotation angles of the HWPs and QWPs shown in Fig.~4.
In the preparation of states $|\psi^{+}\rangle$, $|\psi^{-}\rangle$,
$|\varphi^{+}\rangle$, $|\varphi^{-}\rangle$, the QWP corresponding
to Q$_{2}$ is removed. Details of the angles of wave plates used
to prepare all states are given in Table S1.
\begin{table}[H]\label{table}
\caption{The angles of wave plates used to prepare the states.}
\centering{}%
\begin{tabular}{cccccc}
\toprule
State  & H$_{1}$  & Q$_{1}$  & H$_{3}$  & H$_{2}$  & Q$_{2}$ \tabularnewline
\midrule
$|\psi^{+}\rangle$  & $67.5^{\circ}$  & $45^{\circ}$  & $0^{\circ}$  & $45^{\circ}$  & \textendash{} \tabularnewline
$|\psi^{-}\rangle$  & $22.5^{\circ}$  & $45^{\circ}$  & $0^{\circ}$  & $45^{\circ}$  & \textendash{} \tabularnewline
$|\varphi^{+}\rangle$  & $67.5^{\circ}$  & $45^{\circ}$  & $0^{\circ}$  & \textendash{}  & \textendash{} \tabularnewline
$|\varphi^{-}\rangle$  & $22.5^{\circ}$  & $45^{\circ}$  & $0^{\circ}$  & \textendash{}  & \textendash{} \tabularnewline
$|\Psi(\theta)\rangle$  & $(90-\frac{\theta}{2})^{\circ}$  & $-\theta^{\circ}$  & $45^{\circ}$  & $(90-\frac{\theta}{2})^{\circ}$  & $-\theta^{\circ}$\tabularnewline
\bottomrule
\end{tabular}
\end{table}


\bigskip

\begin{thebibliography}{10}
\bibitem{Baumgratz} Baumgratz, T., Cramer, M., and Plenio, M. B., Quantifying
Coherence, \textit{Phys. Rev. Lett.} \textbf{113}, 140401 (2014).

\bibitem{coherence1} Levi, F. and  Mintert, F., A quantitative theory
of coherent delocalization, \emph{New J. Phys.} \textbf{16}, 033007
(2014).

\bibitem{coherence2} Winter, A. and Yang, D., Operational Resource Theory
of Coherence, \emph{Phys. Rev. Lett.} \textbf{116}, 120404 (2016).

\bibitem{coherence3} Yadin, B., Ma, J., Girolami, D., Gu, M. and Vedral, V.,
Quantum Processes Which Do Not Use Coherence, \emph{Phys. Rev. X}
\textbf{6}, 041028 (2016).

\bibitem{coherence4} Ben Dana, K., García Díaz, M., Mejatty, M. and
 Winter, A., Resource theory of coherence: Beyond states, \emph{Phys.
Rev. A} \textbf{95}, 062327 (2017)

\bibitem{review} Streltsov, A., Adesso, G. and Plenio, M. B., Colloquium:
Quantum coherence as a resource, \textit{Rev. Mod. Phys.} \textbf{89},
041003 (2017).

\bibitem{algorithm1} Hillery, M., Coherence as a resource in decision
problems: The Deutsch-Jozsa algorithm and a variation, \textit{Phys.
Rev. A} \textbf{93}, 012111 (2016).

\bibitem{algorithm2} Shi, H.-L., Liu, S.-Y., Wang, X.-H., Yang, W.-L.,
Yang, Z.-Y. and Fan, H., Coherence depletion in the Grover quantum
search algorithm, \textit{Phys. Rev. A} \textbf{95}, 032307 (2017).

\bibitem{CoherenceComputation} Matera, J. M., Egloff, D., Killoran, N.
and Plenio, M. B., Coherent control of quantum systems as a resource
theory, \emph{Quantum Sci. Technol.} \textbf{1}, 01LT01 (2016).

\bibitem{QKD} Gisin, N., Ribordy, G., Tittel, W. and Zbinden, H., Quantum
cryptography, \textit{Rev. Mod. Phys.} \textbf{74}, 145 (2002).

\bibitem{discrimination2} Napoli, C., Bromley, T. R., Cianciaruso, M.,
Piani, M., Johnston, N. and Adesso, G., Robustness of Coherence: An
Operational and Observable Measure of Quantum Coherence, \textit{Phys.
Rev. Lett.} \textbf{116}, 150502 (2016).

\bibitem{discrimination1} Piani, M., Cianciaruso, M., Bromley, T. R.,
Napoli, C., Johnston, N. and Adesso, G., Robustness of asymmetry and
coherence of quantum states, \textit{Phys. Rev. A} \textbf{93}, 042107
(2016).

\bibitem{metrology1} Giovannetti, V., Lloyd, S. and Maccone, L., Quantum-Enhanced
Measurements: Beating the Standard Quantum Limit, \textit{Science}
\textbf{306}, 1330 (2004).

\bibitem{metrology2} Giovannetti, V., Lloyd, S. and Maccone, L., Advances
in quantum metrology, \textit{Nat. Photonics} \textbf{5}, 222 (2011).

\bibitem{metrology3} Giorda, P. and  Allegra, M., Coherence in quantum
estimation, \textit{J. Phys. A: Math. Theor.} \textbf{51}, 025302
(2018).

\bibitem{asymmetry1} Gour, G. and Spekkens, R. W., The resource theory
of quantum reference frames: manipulations and monotones, \emph{New
J. Phys.} \textbf{10}, 033023 (2008)

\bibitem{asymmetry2} Marvian, I. and Spekkens, R. W., How to quantify
coherence: Distinguishing speakable and unspeakable notions, \emph{Phys.
Rev. A} \textbf{94}, 052324 (2016).

\bibitem{CoherenceEntanglement1} Streltsov, A., Singh, U., Dhar, H. S.,
Bera, M. N. and Adesso, G., Measuring Quantum Coherence with Entanglement,
\emph{Phys. Rev. Lett.} \textbf{115}, 020403 (2015).

\bibitem{CoherenceEntanglement2} Chitambar, E. and Hsieh, M.-H., Relating
the Resource Theories of Entanglement and Quantum Coherence, \emph{Phys.
Rev. Lett}. \textbf{117}, 020402 (2016).

\bibitem{relation1} Tan, K. C., Kwon, H., Park, C.-Y. and Jeong, H.,
Unified view of quantum correlations and quantum coherence, \textit{Phys.
Rev. A} \textbf{94}, 022329 (2016).

\bibitem{conversion2} Chitambar, E., Streltsov, A., Rana, S.,
Bera, M. N., Adesso, G. and Lewenstein, M., Assisted distillation of quantum
coherence, \textit{Phys. Rev. Lett.} \textbf{116}, 070402 (2016).

\bibitem{conversion1} Ma, J., Yadin, B., Girolami, D., Vedral, V. and
Gu, M., Converting coherence to quantum correlations, \textit{Phys.
Rev. Lett.} \textbf{116}, 160407 (2016).

\bibitem{conversion3} Streltsov, A., Rana, S., Bera, M. N. and Lewenstein, M.,
Towards resource theory of coherence in distributed scenarios, \textit{Phys.
Rev. X} \textbf{7}, 011024 (2017).

\bibitem{conversion4} Streltsov, A., Chitambar, E., Rana, S.,
Bera, M. N., Winter, A. and Lewenstein, M., Entanglement and Coherence in
Quantum State Merging, \textit{Phys. Rev. Lett.} \textbf{116}, 240405
(2016).

\bibitem{kangda1} Wu, K.-D., Hou, Z., Zhong, H.-S., Yuan, Y., Xiang, G.-Y.,
Li, C.-F. and Guo, G.-C., Experimentally obtaining maximal coherence
via assisted distillation process, \textit{Optica} \textbf{4}, 4 (2017).

\bibitem{kangda2} Wu, K.-D., Hou, Z., Zhao, Y.-Y., Xiang, G.-Y.,
Li, C.-F., Guo, G.-C., Ma, J., He, Q.-Y., Thompson, J. and Gu, M., Experimental
Cyclic Interconversion between Coherence and Quantum Correlations,
\textit{Phys. Rev. Lett.} \textbf{121}, 050401 (2018).

\bibitem{SingleShotDistillation1} Regula, B., Fang, K., Wang, X. and
Adesso, G., One-Shot Coherence Distillation, \emph{Phys. Rev. Lett.}
\textbf{121}, 010401 (2018).

\bibitem{SingleShotDistillation2} Regula, B., Lami, L. and Streltsov, A.,
Nonasymptotic assisted distillation of quantum coherence, \emph{Phys.
Rev. A} \textbf{98}, 052329 (2018).

\bibitem{SingleShotDistillation3} Vijayan, M. K., Chitambar, E. and
Hsieh, M.-H., One-shot assisted concentration of coherence, \emph{J.
Phys. A.} \textbf{51}, 414001 (2018).

\bibitem{SingleShotDistillation4} Zhao, Q., Liu, Y., Yuan, X., Chitambar, E.
and Winter, A., One-Shot Coherence Distillation: Towards Completing
the Picture, arXiv:1808.01885.

\bibitem{SingleShotDilution} Zhao, Q., Liu, Y., Yuan, X., Chitambar, E.
and Ma, X., One-Shot Coherence Dilution, \emph{Phys. Rev. Lett.} \textbf{120},
070403 (2018).

\bibitem{kangda3} Wu, K.-D., Theurer, T., Xiang, G.-Y., Li, C.-F.,
Guo, G.-C., Plenio, M. B. and Streltsov, A., Quantum coherence and state conversion:
theory and experiment, arXiv:1903.01479.

\bibitem{thermo1} Narasimhachar, V. and Gour, G., Low-temperature thermodynamics
with quantum coherence, \textit{Nat. Commun.} \textbf{6}, 7689 (2015).

\bibitem{thermo2} {Å}berg, J., Catalytic coherence, \textit{Phys.
Rev. Lett.} \textbf{113}, 150402 (2014).

\bibitem{thermo3} Lostaglio, M., Jennings, D. and Rudolph, T., Description
of quantum coherence in thermodynamic processes requires constraints
beyond free energy, \textit{Nat. Commun.} \textbf{6}, 6383 (2015).

\bibitem{thermo4} Korzekwa, K., Lostaglio, M., Oppenheim, J. and
Jennings, D., The extraction of work from quantum coherence, \textit{New
J. Phys.} \textbf{18}, 023045 (2016).

\bibitem{thermo5} Lostaglio, M., Korzekwa, K., Jennings, D. and
Rudolph, T., Quantum coherence, time-translation symmetry, and thermodynamics,
\textit{Phys. Rev. X} \textbf{5}, 021001 (2015).

\bibitem{thermo6} \'{C}wikli\'{n}ski, P., Studzi\'{n}ski, M., Horodecki, M.
and Oppenheim, J., Limitations on the evolution of quantum coherences:
towards fully quantum second laws of thermodynamics, \textit{Phys.
Rev. Lett.} \textbf{115}, 210403 (2015).

\bibitem{thermo7} Rodríguez-Rosario, C. A., Frauenheim, T. and
Aspuru-Guzik, A., Thermodynamics of quantum coherence, arXiv:1308.1245
(2013).

\bibitem{nano} Karlström, O., Linke, H., Karlström, G. and Wacker, A.,
Increasing thermoelectric performance using coherent transport, \textit{Phys.
Rev. B} \textbf{84}, 113415 (2011).

\bibitem{transport1} Herranen, M., Kainulainen, K. and Rahkila, P. M.,
Kinetic transport theory with quantum coherence, \textit{Nucl. Phys.
A} \textbf{820}, 203c (2009).

\bibitem{transport2} Rebentrost, P., Mohseni, M. and Aspuru-Guzik, A.,
Role of quantum coherence and environmental fluctuations in chromophoric
energy transport, \textit{J. Phys. Chem. B} \textbf{113}, 9942 (2009).

\bibitem{biology1} Huelga, S. F. and Plenio, M. B., Vibrations, quanta
and biology, \textit{Contemp. Phys.} \textbf{54}, 181 (2013).

\bibitem{biology2} Lloyd, S., Quantum coherence in biological systems,
\textit{J. Phys.} \textbf{302}, 012037 (2011).

\bibitem{biology3} Plenio, M. B. and Huelga, S. F., Dephasing-assisted
transport: quantum networks and biomolecules, \textit{New J. Phys.}
\textbf{10}, 113019 (2008).

\bibitem{biology4} Lambert, N., Chen, Y.-N., Cheng, Y.-C., Li, C.-M.,
Chen, G.-Y. and Nori, F., Quantum biology, \textit{Nat. Phys.} \textbf{9},
10 (2013).

\bibitem{biology5} Romero, E., Augulis, R., Novoderezhkin, V.I.,
Ferretti, M., Thieme, J., Zigmantas, D. and Van Grondelle, R., Quantum coherence
in photosynthesis for efficient solar-energy conversion, \textit{Nat.
Phys.} \textbf{10}, 676 (2014).

\bibitem{biology6} Huelga, S. F. and Plenio, M. B., Quantum biology:
A vibrant environment, \textit{Nat. Phys.} \textbf{10}, 621 (2014).

\bibitem{wp1} Bera, M. N., Qureshi, T., Siddiqui, M. A. and Pati, A. K.,
Duality of quantum coherence and path distinguishability, \textit{Phys.
Rev. A} \textbf{92}, 012118 (2015).

\bibitem{wp2} Bagan, E., Bergou, J. A., Cottrell, S. S. and Hillery, M.,
Relations between Coherence and Path Information, \textit{Phys. Rev.
Lett.} \textbf{116}, 160406 (2016).

\bibitem{wp3} Yuan, Y., Hou, Z., Zhao, Y.-Y., Zhong, H.-S., Xiang, G.-Y.,
Li, C.-F. and Guo, G.-C., Experimental demonstration of wave-particle
duality relation based on coherence measure. \textit{Opt. Express}
\textbf{26}, 4470 (2018).

\bibitem{QuantifyingEntanglement} Vedral, V., Plenio, M. B., Rippin, M. A.
and Knight, P. L., Quantifying Entanglement, \emph{Phys. Rev. Lett.}
\textbf{78}, 2275 (1997).

\bibitem{ReviewEntanglement} Horodecki, R., Horodecki, P., Horodecki, M.
and Horodecki, K., Quantum entanglement, \emph{Rev. Mod. Phys.} \textbf{81},
865 (2009).

\bibitem{tomography} James, D. F. V., Kwiat, P. G., Munro, W. J. and
 White, A. G., Measurement of qubits, \textit{Phys. Rev. A} \textbf{64},
052312 (2001).

\bibitem{skew} Girolami, D., Observable Measure of Quantum Coherence
in Finite Dimensional Systems, \textit{Phys. Rev. Lett.} \textbf{113},
170401 (2014).

\bibitem{wang} Wang, Y.-T., Tang, J.-S., Wei, Z.-Y., Yu, S., Ke, Z.-J.,
 Xu, X.-Y., Li, C.-F. and Guo, G.-C., Directly Measuring the Degree of
Quantum Coherence using Interference Fringes, \textit{Phys. Rev. Lett.}
\textbf{118}, 020403 (2017).

\bibitem{tong} Zhang, D.-J., Liu, C. L., Yu, X.-D. and Tong, D. M.,
Estimating Coherence Measures from Limited Experimental Data Available,
\textit{Phys. Rev. Lett.} \textbf{120}, 170501 (2018).

\bibitem{njp} Carmeli, C., Heinosaari, T., Maniscalco, S., Schultz, J.
and Toigo, A., Determining quantum coherence with minimal resources,
\textit{New J. Phys.} \textbf{20} 063038 (2018).

\bibitem{coll1} Massar, S. and Popescu, S., Optimal Extraction of Information
from Finite Quantum Ensembles, \textit{Phys. Rev. Lett.} \textbf{74},
1259 (1995).

\bibitem{coll2} Tarrach, R. and Vidal, G., Universality of optimal
measurements, \textit{Phys. Rev. A} \textbf{60}, R3339 (1999).

\bibitem{coll3} Bagan, E., Ballester, M. A., Gill, R. D., Muñoz-Tapia, R.
and Romero-Isart, O., Separable Measurement Estimation of Density Matrices
and its Fidelity Gap with Collective Protocols, \textit{Phys. Rev.
Lett.} \textbf{97}, 130501 (2006).

\bibitem{hou} Hou, Z., Tang, J.-F., Shang, J., Zhu, H., Li, J., Yuan, Y.,
Wu, K.-D., Xiang, G.-Y., Li, C.-F. and Guo, G.-C., Deterministic realization
of collective measurements via photonic quantum walks, \textit{Nat.
Commun.} \textbf{9}, 1414 (2018).

\bibitem{Hu} Ren, H., Lin, A., He, S., Hu, X., Quantitative coherence
witness for finite dimensional states, \textit{Annals of Physics}
\textbf{387}, 281 (2017).

\bibitem{formation} Yuan, X., Zhou, H., Cao, Z. and Ma, X., Intrinsic
randomness as a measure of quantum coherence, \textit{Phys. Rev. A}
\textbf{92}, 022124 (2015).

%\bibitem{maxentropy} K. Bu, U. Singh, S.-M. Fei, A. K. Pati, and
%J. Wu, Maximum Relative Entropy of Coherence: An Operational Coherence
%Measure, \textit{Phys. Rev. Lett.} \textbf{119}, 150405 (2017).
%\bibitem{MeasurementQubits} D. F. V. James, P. G. Kwiat, W. J. Munro,
%and A. G. White, Measurement of qubits, \textit{Phys. Rev. A} \textbf{64},
%052312 (2001).
\bibitem{MLE} Ježek, M., Fiurášek, J. and Hradil, Z., Quantum inference
of states and processes, \textit{Phys. Rev. A} \textbf{68}, 012305 (2003).

\end{thebibliography}
\end{document}